%% file: pipime-paper.tex
\documentclass[11pt]{article}
\usepackage[dvips]{graphicx}
\usepackage{enumitem}
\usepackage{amssymb}
\usepackage{color}
\usepackage{amsmath}
\usepackage{nicefrac}
\usepackage[left]{lineno}
\usepackage{hyperref}
\usepackage{adjustbox}
\usepackage{xspace}
\usepackage{multirow}
\usepackage{alphalph}
\usepackage[dvipsnames]{xcolor}
\usepackage{cite}

\definecolor{blu}{rgb}{0.,0.,1.}
\definecolor{red}{rgb}{1.,0.,0.}
\definecolor{burgundy}{rgb}{0.5, 0.0, 0.13}
\definecolor{crimsonred}{rgb}{0.6, 0.0, 0.0}
\definecolor{persianblue}{rgb}{0.11, 0.22, 0.73}
\definecolor{forestgreen}{rgb}{0.13,0.35,0.13}

\def\geant {\mbox{\textsc{Geant4}}\xspace}

\hypersetup{colorlinks, citecolor=crimsonred, linkcolor=persianblue, urlcolor=crimsonred}

\begin{document}
\centerline{\LARGE EUROPEAN ORGANIZATION FOR NUCLEAR RESEARCH}

% This is to be uncommented after CERN preprint ID is obtained
%
\vspace{10mm} {\flushright{
CERN-EP-2024-224 \\
2 Sep 2024\\
\vspace{4mm}
Revised version:\\7 Nov 2024\\
}}
\vspace{-30mm}

% This is for drafts
%
%\vspace{10mm}
%{\flushright{
%Draft 4\\
%26 August 2024\\
%}}
%\vspace{-15mm}

%%%%%%%%%%%%%%%%%%%%%
%
\vspace{40mm}

\begin{center}
\boldmath
{\bf {\Large\boldmath{First search for $K^+\to\pi^0\pi\mu e$ decays}}}
\unboldmath
\end{center}
\vspace{4mm}
\begin{center}
{\Large The NA62 Collaboration}\\
\end{center}

\begin{abstract}
The first search for the lepton number violating decay $K^+\to\pi^0\pi^-\mu^+e^+$ and lepton flavour violating decays $K^+\to\pi^0\pi^+\mu^-e^+$, $K^+\to\pi^0\pi^+\mu^+e^-$ has been performed using a dataset collected by the NA62 experiment at CERN in 2016--2018. Upper limits of $2.9\times 10^{-10}$, $3.1\times 10^{-10}$ and $5.0\times 10^{-10}$, respectively, are obtained at 90\% CL for the branching ratios of the three decays on the assumption of uniform phase-space distributions.
\end{abstract}

\begin{center}
{\it Accepted for publication in Physics Letters B}
\end{center}

%\begin{linenumbers}

%%%%%%%%%%%%%%%%%%%%%%%%%%%%%%%%%%%%%%%%%%%%%%%%%

\newpage

\section*{Introduction}

In the Standard Model (SM), neutrinos are strictly massless due to the absence of right-handed chiral states. The discovery of neutrino oscillations requires non-zero neutrino masses, making it possible to discriminate experimentally between the Dirac and Majorana neutrino. Strong evidence for the Majorana nature of the neutrino would be provided by the observation of lepton number violating (LNV) processes, including kaon decays~\cite{li00,atre05,atre09,ab18}. Furthermore, lepton flavour violating (LFV) kaon decays are expected in new physics models involving ALPs and $Z'$ particles~\cite{alp2020,review}.

The NA62 experiment at CERN collected a large sample of $K^+$ decays to lepton pairs using dedicated trigger lines in 2016--2018. This dataset has been analysed to establish stringent upper limits on the LNV decays $K^+\to\pi^-(\pi^0)e^+e^+$~\cite{co22}, $K^+\to\pi^-\mu^+\mu^+$~\cite{co19} and $K^+\to\pi^-\mu^+e^+$~\cite{co21}, LFV decays $K^+\to\pi^+\mu^-e^+$ and $\pi^0\to\mu^-e^+$~\cite{co21}, and the $K^+\to\mu^-\nu e^+e^+$ decay violating either LN or LF conservation depending on the flavour of the emitted neutrino~\cite{co23}. The first search for the LNV decay $K^+\to\pi^0\pi^-\mu^+e^+$ and LFV decays $K^+\to\pi^0\pi^+\mu^\pm e^\mp$ performed with the above dataset is reported here.

%%%%%%%%%%%%%%%%%%%%%%%%%%%%%%%%%%%%%%%%%%%%%%%%%

\section{Beam, detector and data sample}
\label{sec:detector}

\begin{figure}[t]
\begin{center}
\resizebox{\textwidth}{!}{\includegraphics{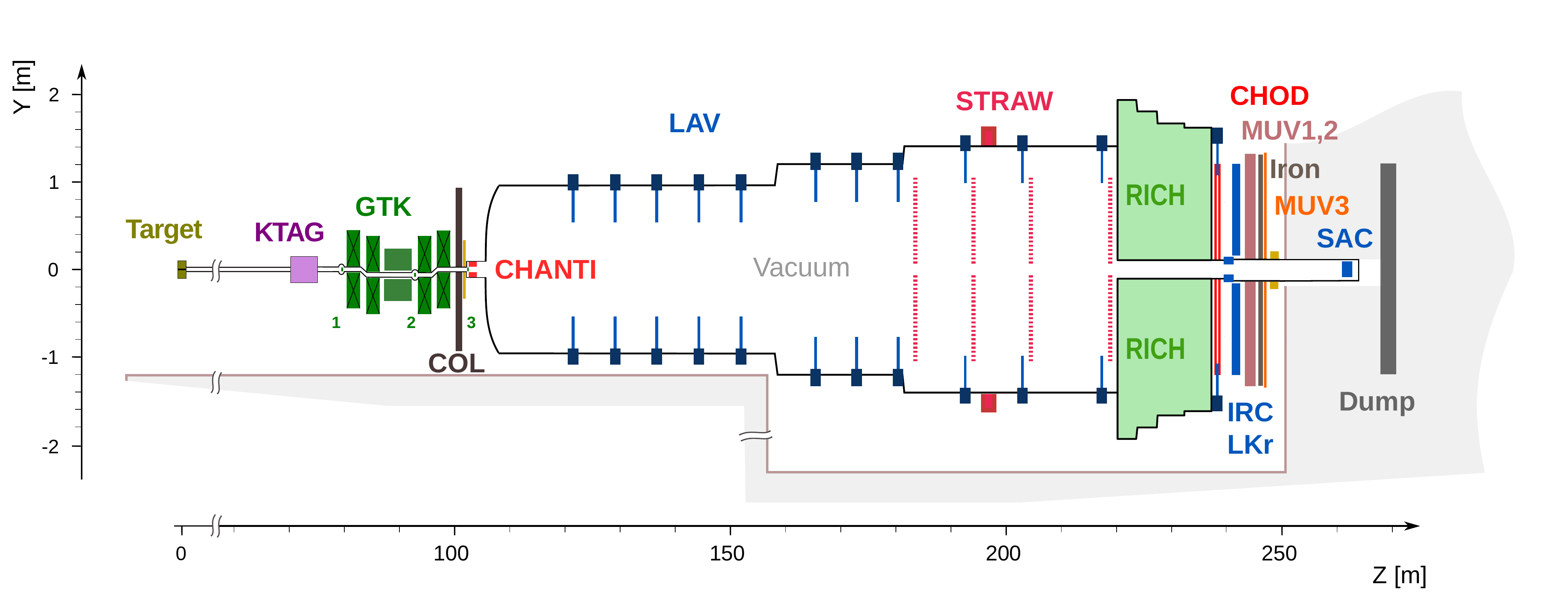}}
\put(-343,68){\scriptsize\color{OliveGreen}\rotatebox{90}{\textbf{\textsf{SCR}}}}
%\put(-325,121){\scriptsize\color{Sepia}\rotatebox{90}{\textbf{\textsf{COL}}}}
\put(-171.5,46){\scriptsize\color{red}{\textbf{\textsf{M}}}}
\put(-319.5,103){\tiny\color{YellowOrange}\rotatebox{90}{\textbf{\textsf{TRIM5}}}}
\end{center}
\vspace{-14mm}
\caption{Schematic side view of the NA62 beamline and detector used in 2018.}
\label{fig:detector}
\end{figure}

The NA62 beamline and detector layout~\cite{na62-detector} used in 2018 is shown schematically in Fig.~\ref{fig:detector}. An unseparated secondary beam of $\pi^+$ (70\%), protons (23\%) and $K^+$ (6\%) is created by directing 400~GeV/$c$ protons extracted from the CERN SPS onto a beryllium target in spills of 4.8~s duration. The beam central momentum is 75~GeV/$c$, with a momentum spread of 1\% (rms).

Beam kaons are tagged with a time resolution of 70~ps by a differential Cherenkov counter (KTAG), which uses nitrogen gas at 1.75~bar pressure contained in a 5~m long vessel as radiator. Beam particle positions, momenta and times (to better than 100~ps resolution) are measured by a silicon pixel spectrometer consisting of three stations (GTK1,2,3) and four dipole magnets forming an achromat. A toroidal muon sweeper (scraper, SCR) is installed between GTK1 and GTK2. A 1.2~m thick steel collimator (COL) with a $76\times40$~mm$^2$ central aperture and $1.7\times1.8$~m$^2$ outer dimensions is placed upstream of GTK3 to absorb hadrons from upstream $K^+$ decays; a variable-aperture collimator of \mbox{$0.15\times0.15$~m$^2$} outer dimensions was used up to early 2018. Inelastic interactions of beam particles in GTK3 are detected by an array of scintillator hodoscopes (CHANTI). A dipole magnet (TRIM5) providing a 90~MeV/$c$ horizontal momentum kick is located in front of GTK3. The beam is delivered into a vacuum tank evacuated to a pressure of $10^{-6}$~mbar, which contains a 75~m long fiducial volume (FV) starting 2.6~m downstream of GTK3. The beam angular spread at the FV entrance is 0.11~mrad (rms) in both horizontal and vertical planes. The probability of beam kaon decay in the FV is 12.5\%. Downstream of the FV, undecayed beam particles continue their path in vacuum.

Momenta of charged particles produced in $K^+$ decays in the FV are measured by a magnetic spectrometer (STRAW) located in the vacuum tank downstream of the FV. The spectrometer consists of four tracking chambers made of straw tubes, and a dipole magnet (M) located between the second and third chambers that provides a horizontal momentum kick of 270~MeV/$c$ in a direction opposite to that produced by TRIM5. The momentum resolution is $\sigma_p/p = (0.30\oplus 0.005\cdot p)\%$, with the momentum $p$ expressed in GeV/$c$.

% RICH threshold for pions quoted according to the detector paper.
% Another estimate: 90 * m_pi = 12.6 GeV/c.
%
A ring-imaging Cherenkov detector (RICH) consisting of a 17.5~m long vessel filled with neon at atmospheric pressure (with a Cherenkov threshold of 12.5~GeV/$c$ for pions) provides particle identification, charged particle time measurements with a typical resolution of 70~ps, and the trigger time.
%The RICH optical system is optimised to collect light emitted by positively charged particles, exploiting their deflection by the STRAW dipole magnet.
Two scintillator hodoscopes (CHOD), which include a matrix of tiles and two planes of slabs arranged in four quadrants located downstream of the RICH, provide trigger signals and time measurements with 200~ps precision.

\newpage

%%%%%%%%%%%%%%%%%%%%%%%%%%%%%%%%%%

A $27X_0$ thick quasi-homogeneous liquid-krypton (LKr) electromagnetic calorimeter is used for particle identification and photon detection. The calorimeter has an active volume of 7~m$^3$, segmented in the transverse direction into 13248 projective cells of $2\times 2$~cm$^2$ size, and provides energy resolution $\sigma_E/E=(4.8/\sqrt{E}\oplus11/E\oplus0.9)\%$, with $E$ expressed in GeV. To achieve hermetic acceptance for photons emitted in $K^+$ decays in the FV at angles up to 50~mrad from the beam axis, the LKr calorimeter is complemented by annular lead glass detectors (LAV) installed in 12~positions inside and downstream of the vacuum tank, and two lead/scintillator sampling calorimeters (IRC, SAC) located close to the beam axis. An iron/scintillator sampling hadronic calorimeter formed of two modules (MUV1,2) and a muon detector (MUV3) consisting of 148~scintillator tiles located behind an 80~cm thick iron wall are used for particle identification.

%
% Data sample used: 16 [~60k bursts with ee trigger] +
% 17 [301k bursts per Twiki] + 18 [525k bursts per Twiki] = 885k bursts
%

The data sample analysed is obtained from $8.9\times 10^5$ SPS spills recorded in 2016--2018. The typical beam intensity was increased during the data collection time from \mbox{$1.3\times 10^{12}$} to \mbox{$2.2\times 10^{12}$} protons per spill. The latter value corresponds to a 500~MHz mean instantaneous beam particle rate at the FV entrance, and a 3.7~MHz mean $K^+$ decay rate in the FV. The main NA62 trigger line is designed for the collection of the very rare $K^+\to\pi^+\nu\bar\nu$ decay~\cite{pinn}. The present analysis is based on the dedicated multi-track (MT), electron multi-track ($e$MT) and muon multi-track ($\mu$MT) trigger lines operating concurrently with the main trigger line~\cite{am19,na62-trigger}, and downscaled typically by factors $D_{\rm MT}=100$, $D_{e\rm MT}=8$ and $D_{\mu\rm MT}=8$. The downscaling factors were varied during data-taking to accommodate the changes in the beam intensity.

The low-level (L0) hardware trigger in all three trigger lines is based on RICH signal multiplicity and coincidence of signals in two opposite CHOD quadrants. The $\mu$MT ($e$MT) line additionally requires a minimum energy deposit of 10~(20)~GeV in the LKr calorimeter. The $\mu$MT line also requires a signal in an outer MUV3 detector tile (i.e. one of the 140 tiles not adjacent to the beam pipe). The high-level (L1) software trigger involves beam $K^+$ identification by the KTAG, reconstruction of a negatively-charged STRAW track and, only for the $\mu$MT trigger line, fewer than three in-time signals in LAV detectors 2--11. For signal-like samples characterised by LKr energy deposit well above 20~GeV, the measured inefficiencies of the CHOD (STRAW) trigger conditions are typically at the 1\%~(5\%) level, while those of the RICH, MUV3, KTAG and LKr conditions are of ${\cal O}(10^{-3})$.

Monte Carlo simulations of particle interactions with the detector and its response are performed using a software package based on the~\geant toolkit~\cite{geant4}. In addition, accidental activity is simulated and the response of the trigger lines is emulated.

%%%%%%%%%%%%%%%%%%%%%%%%%%%%%%%%%%%%%%%%%%%%%%%%%%%

\section{Event selection}
\label{sec:selection}

The rates of the possible signal decays $K^+\to\pi^0\pi\mu e$ (denoted $K_{\pi\pi\mu e}$ below) are measured with respect to the rate of the normalisation decay $K^+\to\pi^+e^+e^-$ (denoted $K_{\pi ee}$ below), providing partial cancellation of detector and trigger inefficiencies. The $K_{\pi\pi\mu e}$ decay candidates are collected with the MT, $e$MT and $\mu$MT trigger lines, while the $K_{\pi ee}$ decay candidates are collected with the MT and $e$MT lines only. The following principal selection criteria are common for the $K_{\pi\pi\mu e}$ and $K_{\pi ee}$ candidates.
\begin{itemize}
\item Three-track vertices are reconstructed by extrapolating STRAW tracks into the FV, taking into account the measured residual magnetic field in the vacuum tank, and selecting triplets of tracks consistent with originating from the same point. Exactly one vertex should be present in the event. The total charge of the three tracks should be $q=+1$. The longitudinal position of the vertex, $z_{\rm vtx}$, should be within the FV. The momenta of the tracks forming the vertex should exceed 6~GeV/$c$. Track trajectories through the STRAW chambers and extrapolated positions in the CHOD and LKr calorimeter front planes should be within the respective geometrical acceptances. Tracks should be separated from each other by at least 15~mm in each STRAW chamber plane to suppress photon conversions, and at least 200~mm in the LKr front plane  to reduce the effects of shower overlaps.
\item Track times are initially defined using the CHOD information. The vertex time is initially evaluated as the average of the track CHOD times. Signals in the RICH geometrically compatible with the tracks, and within 3~ns of the vertex time, are used to evaluate track RICH times. Track and vertex time estimates are then refined using the RICH information. Each track is required to be within 2.5~ns of the trigger time.
\item No signals are allowed in the LAV detectors downstream of the reconstructed vertex position within 4~ns of the vertex time. This condition suppresses backgrounds with photons in the final state. Most importantly, the $K^+\to\pi^+\pi^0\pi^0_{\rm D}$, $\pi^0_{\rm D}\to\gamma e^+e^-$ background to the $K_{\pi\pi\mu e}$ decays is reduced by one order of magnitude.
\item Particle identification is based on the ratio $E/p$ of the energy deposited in the LKr calorimeter (within 50~mm of the track impact point and within 10~ns of the vertex time) to the momentum measured by the spectrometer. Pion ($\pi^\pm$), muon ($\mu^\pm$) and electron ($e^\pm$) candidates are required to have $E/p<0.85$, $E/p<0.2$ and $0.9<E/p<1.1$, respectively. An associated MUV3 signal within 5~ns of the vertex time is required for muon candidates, while no such MUV3 signals are allowed for pion candidates.
\end{itemize}
%
%%%%%%%%%%%%%%%%%%%%
%
The $K_{\pi ee}$ selection, identical to that of Ref.~\cite{co23}, includes the following additional criteria.
\begin{itemize}
\item The tracks forming the vertex should be identified as $\pi^+e^+e^-$, according to the above particle identification criteria.
\item The total momentum of the three tracks, $p_{\rm vtx}$, should satisfy the condition $|p_{\rm vtx}-p_{\rm beam}|<2~{\rm GeV}/c$, where $p_{\rm beam}$ is the peak value of the beam momentum. The total transverse momentum with respect to the beam axis should be below 30~MeV/$c$. The quantity $p_{\rm beam}$ and the beam axis direction, averaged over a few hours, are monitored throughout the data taking with fully reconstructed $K^+\to\pi^+\pi^+\pi^-$ decays.
\item The reconstructed $\pi^+e^+e^-$ mass, $m_{\pi ee}$, should be in the normalisation region defined as 470--505~MeV/$c^2$, accounting for the mass resolution of 1.7~MeV/$c^2$ and the radiative tail. The reconstructed $e^+e^-$ mass should be $m_{ee}>140~{\rm MeV}/c^2$ to suppress backgrounds from the $K^+\to\pi^+\pi^0$ decay followed by $\pi^0_{\rm D}\to e^+e^-\gamma$, $\pi^0_{\rm DD}\to e^+e^-e^+e^-$ and $\pi^0\to e^+e^-$ decays. This leads to a relative acceptance reduction of 27\%.
\end{itemize}
%
%%%%%%%%%%%%%%%%%%%%
%
The following conditions are used to select the $K_{\pi\pi\mu e}$ candidates.
\begin{itemize}
\item The tracks forming the vertex should be identified as $\pi\mu e$, according to the above particle identification criteria. Track charges should correspond to one of the three signal decays, denoted unambiguously as the $\pi^-$, $\mu^-$ and $e^-$ modes below.
\item The $\pi^0$ meson is reconstructed via the $\pi^0\to\gamma\gamma$ decay. Exactly two photon candidates are required, defined as reconstructed LKr energy deposit clusters within the geometrical acceptance, with energy above 2~GeV, within 5~ns of the vertex time, and separated by at least 150~mm from each other and from each track impact point in the LKr calorimeter front plane.
\item The longitudinal coordinate of the ``neutral vertex'' is defined
assuming a prompt $\pi^0\to\gamma\gamma$ decay: $z_{\rm N} = z_{\rm LKr} - D_{12}\sqrt{E_1E_2}/m_{\pi^0}$. Here $D_{12}$ is the distance between the photon candidates in the LKr transverse plane at the $z$ coordinate $z_{\rm LKr}$; $E_{1,2}$ are the photon candidate energies; $m_{\pi^0}$ is the nominal $\pi^0$ mass~\cite{pdg}.
\item Consistency of the three-track and neutral vertices is required: $|z_{\rm vtx}-z_{\rm N}|<8$~m. Vertex position resolutions evaluated with simulations are $\delta z_{\rm vtx}=0.25$~m and $\delta z_{\rm N}=1.8$~m.
\item Photon momenta are computed using photon candidate energies and positions in the LKr calorimeter transverse plane, assuming emission at the three-track vertex. The $\pi^0$ momentum is computed as the sum of photon momenta, and the $\pi^0$ energy is computed using the $\pi^0$ mass. 
\item The total final-state momentum, $p_{\pi\pi\mu e}$, should be consistent with the beam momentum: the difference, $\Delta p=p_{\pi\pi\mu e}-p_{\rm beam}$, should satisfy the condition $|\Delta p|<3~{\rm GeV}/c$. The total transverse momentum of the final-state particles with respect to the beam axis should be $p_T<30~{\rm MeV}/c$.
\item The reconstructed $\pi^0\pi\mu e$ mass, $m_{\pi\pi\mu e}$, should be in the signal region 486--502~MeV/$c^2$, which accounts for the mass resolution of 1.3~MeV/$c^2$ and non-gaussian tails.
\end{itemize}

%%%%%%%%%%%%%%%%%%%%%%%%%%%%%

\begin{figure}[p]
\begin{center}
\resizebox{0.5\textwidth}{!}{\includegraphics{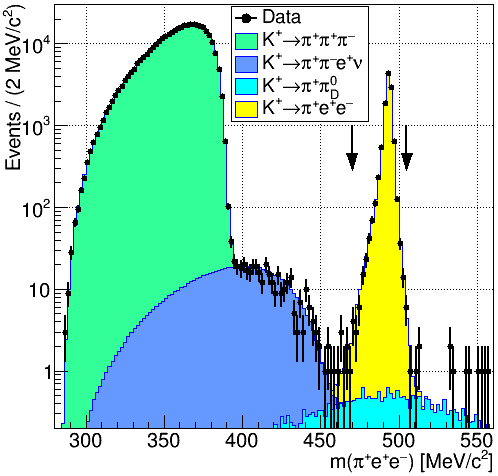}}
\end{center}
\vspace{-16mm}
\caption{Reconstructed $m_{\pi ee}$ spectra for data (with statistical errors) and simulated samples obtained with the $K_{\pi ee}$ selection. The normalisation region is shown with vertical arrows. The data events are not weighted; see Section~\ref{sec:piee} for details.}
\label{fig:mpiee}
\vspace{-3mm}
\end{figure}

%%%%%%%%%%%%%%%%%%%%%%%%%%%%%%%%%%%%%%%%%%%%%%%%%%%

\begin{table}[p]
\setlength{\tabcolsep}{4pt} % reduced column width
\caption{Sources of background to $K_{\pi\pi\mu e}$ decays, decay branching ratios (or their products), and misidentification processes leading to each of the three signal final states. In some cases of misidentification, a $\pi^\pm\to\mu^\pm\nu$ decay in flight is additionally required. Cases with no misidentification are classified as ``direct'' (a signal final state is produced directly), and ``$\pi^\pm$ DIF'' (with a $\pi^\pm\to\mu^\pm\nu$ decay in flight leading to the signal final state).}
\begin{center}
\vspace{-9mm}
\begin{tabular}{l|c|ccc}
\hline
Background source & ${\cal B}$ & $\pi^0\pi^-\mu^+e^+$ & $\pi^0\pi^+\mu^-e^+$ & $\pi^0\pi^+\mu^+e^-$ \\
\hline
$K^+\to\pi^+\pi^0\pi^0_{\rm D}$ & $4.1\times 10^{-4}$ &
$e^-\to\pi^-$ & $e^-\to\mu^-$ & $e^+\!\to\!\pi^+$ or $e^+\!\to\!\mu^+$ \\
$K^+\to\pi^+\pi^0_{\rm D}\gamma$ ($E_\gamma^*\!>\!10$~MeV) &
$7.8\times 10^{-6}$ &
$e^-\to\pi^-$ & $e^-\to\mu^-$ & $e^+\!\to\!\pi^+$ or $e^+\!\to\!\mu^+$ \\
$K^+\to\pi^+\pi^0 e^+e^-$ & $4.2\times 10^{-6}$ &
$e^-\to\pi^-$ & $e^-\to\mu^-$ & $e^+\!\to\!\pi^+$ or $e^+\!\to\!\mu^+$ \\
$K^+\to\pi^0_{\rm D}\mu^+\nu\gamma$ ($E_\gamma^*\!>\!10$~MeV) &
$7.4\times 10^{-7}$ &
$e^-\to\pi^-$ & -- & $e^+\to\pi^+$ \\
$K^+\to\pi^0\pi^0_{\rm D}\mu^+\nu$ & $7.9\times 10^{-8}$ &
$e^-\to\pi^-$ & -- & $e^+\to\pi^+$ \\
$K^+\to\pi^+\pi^0\pi^0$, $\pi^0\to e^+e^-$ & $2.6\times 10^{-9}$ &
$e^-\to\pi^-$ & $e^-\to\mu^-$ & $e^+\!\to\!\pi^+$ or $e^+\!\to\!\mu^+$ \\
\hline
$K^+\to\pi^+\pi^0 + K^+\to\pi^+\pi^+\pi^-$ & $1.1\times 10^{-2}$ &
$\pi^+\to e^+$ & $\pi^+\to e^+$ & $\pi^-\to e^-$ \\
$K^+\to\pi^+\pi^0\pi^0 + K^+\to\pi^+\pi^+\pi^-$ & $1.0\times 10^{-3}$ &
$\pi^+\to e^+$ & $\pi^+\to e^+$ & $\pi^-\to e^-$ \\
$K^+\to\pi^0e^+\nu+K^+\to\pi^+\pi^+\pi^-$ & $2.8\times 10^{-3}$ &
$\pi^+$ DIF & $\pi^-$ DIF & $\pi^-\to e^-$ \\
$K^+\to\pi^0\mu^+\nu+K^+\to\pi^+\pi^+\pi^-$ & $1.8\times 10^{-3}$ &
$\pi^+\to e^+$ & $\pi^+\to e^+$ & $\pi^-\to e^-$ \\
$K^+\to\pi^+\pi^0_{\rm D}+K^+\to\mu^+\nu$ & $1.5\times 10^{-3}$ &
-- & -- & $e^+\to\gamma$ \\
$K^+\to\pi^+\pi^0_{\rm D}+K^+\to\pi^+\pi^0$ & $5.0\times 10^{-4}$ &
$e^-\to\pi^-$ & $e^-\to\mu^-$ & $\pi^+$ DIF \\
$K^+\to\pi^+\pi^0\pi^0_{\rm D}+K^+\to\mu^+\nu$ & $2.6\times 10^{-4}$ &
$e^-\to\pi^-$ & $e^-\to\mu^-$ & direct \\
$K^+\to\pi^+\pi^0\pi^0_{\rm D}+K^+\to\pi^+\pi^0$ & $8.3\times 10^{-5}$ &
$e^-\to\pi^-$ & $e^-\to\mu^-$ & $\pi^+$ DIF \\
$K^+\to\pi^0\mu^+\nu+K^+\to\pi^+\pi^0_{\rm D}$ & $8.0\times 10^{-5}$ &
$e^-\to\pi^-$ & $e^-\to\mu^-$ & direct \\
$K^+\to\pi^0_{\rm D}\mu^+\nu+K^+\to\pi^+\pi^0$ & $8.0\times 10^{-5}$ &
$e^-\to\pi^-$ & $e^-\to\mu^-$ & direct \\
$K^+\to\pi^+\pi^0\pi^0+K^+\to\pi^+\pi^0_{\rm D}$ & $4.2\times 10^{-5}$ &
$e^-\to\pi^-$ & $e^-\to\mu^-$ & $\pi^+$ DIF \\
$K^+\to\pi^0_{\rm D}e^+\nu+K^+\to\pi^0\mu^+\nu$ & $2.0\times 10^{-5}$ &
$e^-\to\pi^-$ & -- & $e^+\to\pi^+$ \\
$K^+\to\pi^0e^+\nu+K^+\to\pi^0_{\rm D}\mu^+\nu$ & $2.0\times 10^{-5}$ &
$e^-\to\pi^-$ & -- & $e^+\to\pi^+$ \\
$K^+\to\pi^0\mu^+\nu+K^+\to\pi^+\pi^0\pi^0_{\rm D}$ & $1.4\times 10^{-5}$ &
$e^-\to\pi^-$ & $e^-\to\mu^-$ & direct \\
$K^+\to\pi^0\mu^+\nu+K^+\to\pi^0_{\rm D}\mu^+\nu$ & $1.3\times 10^{-5}$ &
$e^-\to\pi^-$ & -- & $e^+\to\pi^+$ \\
$K^+\to\pi^0_{\rm D}\mu^+\nu+K^+\to\pi^+\pi^0\pi^0$ & $6.8\times 10^{-6}$ &
$e^-\to\pi^-$ & $e^-\to\mu^-$ & direct \\
$K^+\to\pi^+\pi^-e^+\nu+K^+\to\pi^0\mu^+\nu$ & $1.4\times 10^{-6}$ &
direct & $\pi^-$ DIF & $\pi^-\to e^-$ \\
$K^+\to\pi^0\pi^0e^+\nu+K^+\to\pi^+\pi^+\pi^-$ & $1.4\times 10^{-6}$ &
$\pi^+$ DIF & $\pi^-$ DIF & $\pi^-\to e^-$ \\
\hline
\end{tabular}
\end{center}
\vspace{-12mm}
\label{tab:bkg1}
\end{table}

%%%%%%%%%%%%%%%%%%%%%%%%%%%%%%%%%%%%%%%

\boldmath
\section{Evaluation of the effective number of $K^+$ decays}
\unboldmath
\label{sec:piee}

The reconstructed $m_{\pi ee}$ spectra obtained using the $K_{\pi ee}$ selection for the data, simulated signal and background samples are displayed in Fig.~\ref{fig:mpiee}. Below the $m_{\pi ee}$ normalisation region, the background is mainly due to $K^+\to\pi^+\pi^+\pi^-$ decays with two pions ($\pi^\pm$) misidentified as electrons ($e^\pm$), and $K^+\to\pi^+\pi^-e^+\nu$ decays with a pion ($\pi^-$) misidentified as an electron ($e^-$). In the $m_{\pi ee}$ normalisation region, 10975 decay candidates are observed in the data, with the background coming mainly from the $K^+\to\pi^+\pi^0_{\rm D}$, $\pi^0_{\rm D}\to\gamma e^+e^-$ decay chain. This background is suppressed by the selection condition $m_{ee}>140~{\rm MeV}/c^2$, and contributes via double particle misidentification ($\pi^+\to e^+$ and $e^+\to\pi^+$).

To account for the fact that the $\mu$MT trigger line is used to collect $K_{\pi\pi\mu e}$ candidates only, while the $e$MT and MT lines are used to collect both $K_{\pi\pi\mu e}$ and $K_{\pi ee}$ candidates, a weight determined by the trigger downscaling factors is applied to each $K_{\pi ee}$ candidate in the data sample to evaluate the effective number of $K_{\pi ee}$ candidates for normalisation:
\begin{displaymath}
w = \frac
{1-\left(1-\frac{1}{D_{e{\rm MT}}}\right)
\left(1-\frac{1}{D_{\mu{\rm MT}}}\right)
\left(1-\frac{1}{D_{{\rm MT}}}\right)}
{1-\left(1-\frac{1}{D_{e{\rm MT}}}\right)
\left(1-\frac{1}{D_{{\rm MT}}}\right)}.
\end{displaymath}

\newpage

The weight quantifies the enhancement of the $K^+$ flux provided by the $\mu$MT trigger line collecting $K_{\pi\pi\mu e}$ candidates, and varies between 1.0 and 2.9 depending on the trigger configuration. Each factor $(1\!-\!1/D)$ represents the probability for an event not to be collected by a trigger line due to the downscaling applied.

The effective number of $K^+$ decays in the FV is computed as
\begin{displaymath}
N_K = \frac{(1-f)\cdot N_{\pi ee}}{{\cal B}_{\pi ee}\cdot A_{\pi ee}} = (1.97 \pm 0.02_{\rm stat} \pm 0.02_{\rm syst} \pm 0.06_{\rm ext})\times 10^{12},
\end{displaymath}
where $N_{\pi ee}=21401$ is the number of weighted $K_{\pi ee}$ candidates in the data sample; ${\cal B}_{\pi ee}=(3.00\pm0.09)\times 10^{-7}$ is the $K_{\pi ee}$ branching ratio~\cite{pdg}; $A_{\pi ee}=(3.62\pm0.03_{\rm syst})\times 10^{-2}$ is the selection acceptance evaluated with simulations; and $f=1.0\times 10^{-3}$ is the relative background contamination evaluated with simulations. The uncertainty in $A_{\pi ee}$ is estimated by varying the selection criteria. The statistical uncertainty in $N_K$ is due to the finite number of $K_{\pi ee}$ candidates, the systematic uncertainty is due to $A_{\pi ee}$, and the external uncertainty is due to ${\cal B}_{\pi ee}$. The normalisation sample and the $N_K$ value are identical to those of Ref.~\cite{co23}.

%%%%%%%%%%%%%%%%%%%%%%%%%%%%%%%%%%%%%%%

\begin{figure}[p]
\vspace{-5mm}
\begin{center}
\resizebox{0.48\textwidth}{!}{\includegraphics{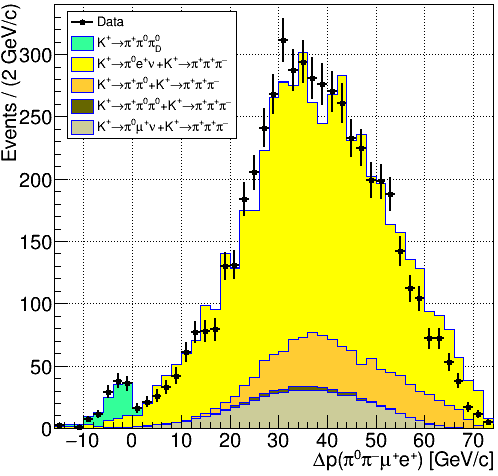}}%
\resizebox{0.48\textwidth}{!}{\includegraphics{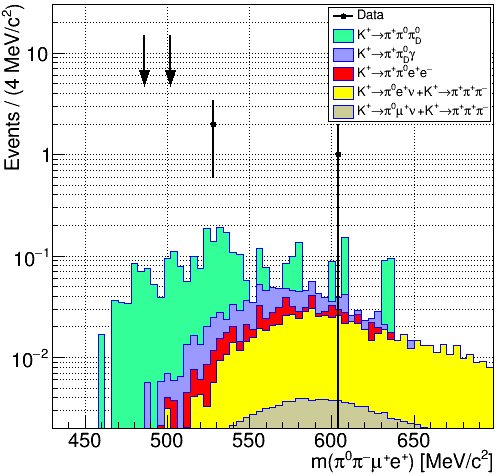}}%

\resizebox{0.48\textwidth}{!}{\includegraphics{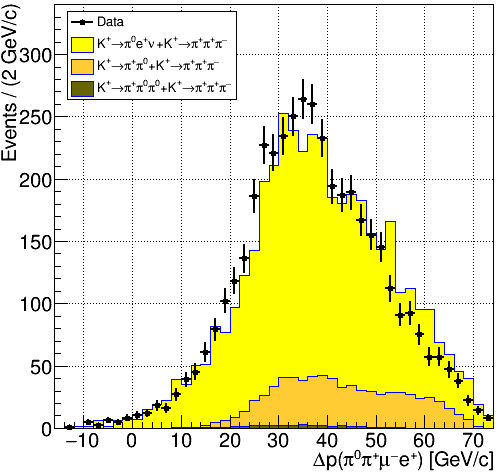}}%
\resizebox{0.48\textwidth}{!}{\includegraphics{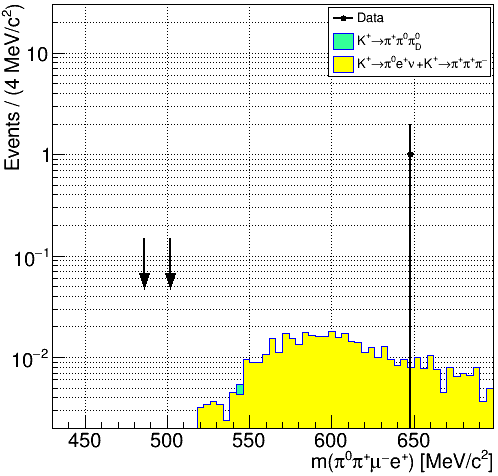}}%

\resizebox{0.48\textwidth}{!}{\includegraphics{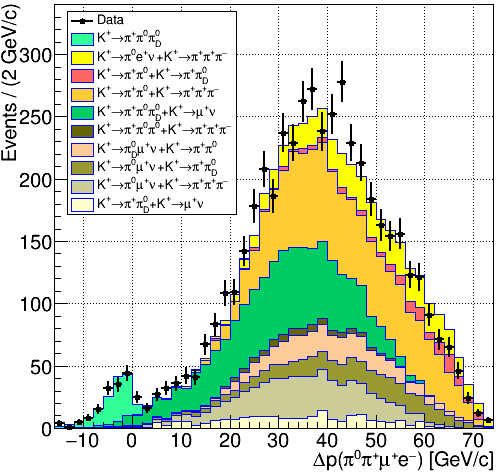}}%
\resizebox{0.48\textwidth}{!}{\includegraphics{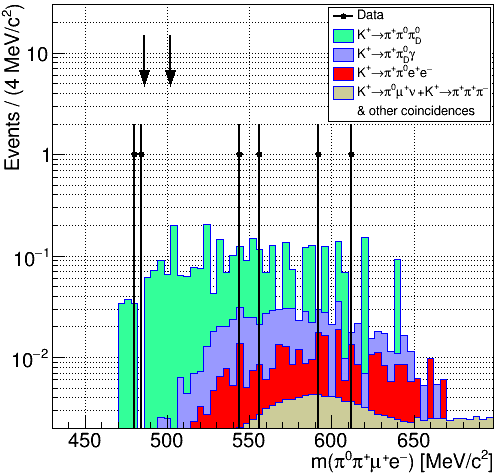}}%
\end{center}
\vspace{-15.5mm}
\caption{Left: momentum difference ($\Delta p$) spectra for data and simulated samples obtained using a loose $K_{\pi\pi\mu e}$ selection without the $\Delta p$, $p_T$ and LAV veto conditions, and with the $m_{\pi\pi\mu e}$ condition inverted. Right: mass ($m_{\pi\pi\mu e}$) spectra for data and simulated samples for the full selection; the signal $m_{\pi\pi\mu e}$ regions, shown with arrows, are not masked.
Top row: $K^+\to\pi^0\pi^-\mu^+e^+$ mode;
middle row: $K^+\to\pi^0\pi^+\mu^-e^+$ mode;
bottom row: $K^+\to\pi^0\pi^+\mu^+e^-$ mode.}
\vspace{-1mm}
\label{fig:signal}
\end{figure}

%%%%%%%%%%%%%%%%%%%%%%%%%%%%%%%%%%%%%%%

\vspace{-1.2mm}
\boldmath
\section{Evaluation of the background to $K_{\pi\pi\mu e}$ decays}
\unboldmath
\label{sec:background}
\vspace{-0.3mm}

Background to the $K_{\pi\pi\mu e}$ decays is evaluated with simulations. The signal $m_{\pi\pi\mu e}$ regions (486--502~MeV/$c^2$) are kept masked for the data until the background estimates are finalised. Backgrounds from single $K^+$ decays and from coincidences of pairs of $K^+$ decays are considered. Simulated samples of pairs of decays occurring simultaneously are used in the latter case. Background is primarily due to particle misidentification. A dedicated data-driven particle identification model~\cite{co22} is employed in the simulations: $\pi^\pm$ and $e^\pm$ (mis)identification probabilities measured with $K^+\to\pi^+\pi^+\pi^-$ and $K^+\to\pi^0e^+\nu$ decay samples are applied as weights to simulated events. The misidentification probabilities are found to be ${\cal O}(10^{-2})$ for $\pi^\pm\to e^\pm$ and $e^\pm\to\pi^\pm$ cases, and ${\cal O}(10^{-6})$ for the $e^\pm\to\mu^\pm$ case, and depend on momentum~\cite{co22,co23}. The data-driven approach avoids \geant-based modelling of the quantity $E/p$ sensitive to simulation of hadronic showers, and increases the effective simulated statistics. Background sources, their branching ratios~\cite{pdg} used for normalisation, and misidentification processes involved in each case are listed in Table~\ref{tab:bkg1}.

Estimation of the background from single $K^+$ decays with the above method was validated to a 20\% precision in a dedicated study~\cite{co22}. To validate the description of the background from coincidences, a loose selection is used, obtained from the signal selection by removing the $\Delta p$, $p_T$ and LAV veto conditions, and inverting the $m_{\pi\pi\mu e}$ condition. Momentum difference ($\Delta p$) spectra for data and simulated samples obtained using the loose selection for each of the three $K_{\pi\pi\mu e}$ modes are shown in Fig.~\ref{fig:signal}~(left). The regions of large $\Delta p$ are populated exclusively by backgrounds from coincidences. The region $\Delta p>10$~GeV/$c$ for the $\mu^-$ mode (with the simplest background structure among the three modes) is used to normalise the backgrounds from coincidences, accounting for the mean probability of coincidence of two decays in the selection time window. The maximum deviation from unity of the ratios of data and simulated $\Delta p$ spectra (Fig.~\ref{fig:signal}, left) and $p_T$ spectra (not shown) is found to be 40\%.

Mass spectra for data and simulated background samples obtained with the full $K_{\pi\pi\mu e}$ selection are shown in Fig.~\ref{fig:signal}~(right). For the $\pi^-$ and $e^-$ modes, background in the signal region is dominated by $K^+\to\pi^+\pi^0\pi^0_{\rm D}$ decays with an undetected soft photon from the $\pi^0_{\rm D}\to\gamma e^+e^-$ decay, a $\pi^+\to\mu^+\nu$ decay in flight, and $e^-\to\pi^-$ ($e^+\to\pi^+$) misidentification in the $\pi^-$ ($e^-$) case. The missing photon and neutrino lead to a negative mean $\Delta p$ value, as seen for the loose selection in Fig.~\ref{fig:signal}~(left). For the $\mu^-$ mode, the $K^+\to\pi^+\pi^0\pi^0_{\rm D}$ contribution is suppressed by the low $e^-\to\mu^-$ misidentification probability. The background is smaller than for the other modes, and is dominated by coincidences of a $K^+\to\pi^0e^+\nu$ and a $K^+\to\pi^+\pi^+\pi^-$ decay, with a $\pi^-\to\mu^-\nu$ decay in flight and a $\pi^+$ not reconstructed in the spectrometer.

\newpage

Background estimates in the sidebands of the $m_{\pi\pi\mu e}$ spectra, i.e. outside the signal regions in Fig.~\ref{fig:signal}~(right), are compared to the observed numbers of data events for the full signal selection in Table~\ref{tab:bkg2}. A signal selection without the LAV veto condition is considered as a cross-check; background estimates for this selection are also compared to the data in Table~\ref{tab:bkg2}. In all cases, data and background estimates in the $m_{\pi\pi\mu e}$ sidebands agree within statistical uncertainties, which further validates the background evaluation procedure.

The final background estimates in the signal region are
\begin{eqnarray}
K^+\to\pi^0\pi^-\mu^+e^+: & N_{\rm bkg} = & \!\!\!0.33\pm0.07,\nonumber \\
K^+\to\pi^0\pi^+\mu^-e^+: & N_{\rm bkg} = & \!\!\!0.004\pm0.003,\nonumber \\
K^+\to\pi^0\pi^+\mu^+e^-: & N_{\rm bkg} = & \!\!\!0.29\pm0.07. \nonumber
\end{eqnarray}
The uncertainties quoted above include a statistical component due to the limited size of simulated samples, and systematic components obtained by background studies as discussed above.

\begin{table}[t]
\caption{Background estimates in the sidebands of the $m_{\pi\pi\mu e}$ spectra with their statistical uncertainties, compared to the numbers of events observed in the data sample.}
\vspace{-10mm}
\begin{center}
\begin{tabular}{l|rcl|c|rcl|c}
\hline
& \multicolumn{4}{c|}{Full signal selection} &
\multicolumn{4}{c}{Selection without the LAV veto} \\
\cline{2-9}
Mode & \multicolumn{3}{c|}{Estimate} & Data & \multicolumn{3}{c|}{Estimate} & Data \\
\hline
$K^+\to\pi^0\pi^-\mu^+e^+$ &
3.26 & $\!\!\pm\!\!$ & 0.30 & 3 & 23.7 & $\!\!\pm\!\!$ & 0.8 & 25 \\
$K^+\to\pi^0\pi^+\mu^-e^+$ &
0.46 & $\!\!\pm\!\!$ & 0.14 & 1 & 0.85 & $\!\!\pm\!\!$ & 0.24 & 1 \\
$K^+\to\pi^0\pi^+\mu^+e^-$ &
3.22 & $\!\!\pm\!\!$ & 0.28 & 6 & 31.7 & $\!\!\pm\!\!$ & 1.0 & 36 \\
\hline
\end{tabular}
\end{center}
\vspace{-13mm}
\label{tab:bkg2}
\end{table}

%%%%%%%%%%%%%%%%%%%%%%%

\section{Results}

Signal acceptances with their statistical uncertainties are evaluated with simulations, assuming uniform phase-space distributions:
\begin{eqnarray}
A(K^+\to\pi^0\pi^-\mu^+e^+) & = & (4.07\pm0.06)\times 10^{-3}, \nonumber \\
A(K^+\to\pi^0\pi^+\mu^-e^+) & = & (3.76\pm0.06)\times 10^{-3}, \nonumber \\
A(K^+\to\pi^0\pi^+\mu^+e^-) & = & (2.37\pm0.05)\times 10^{-3}. \nonumber
\end{eqnarray}
The uncertainties quoted above are dominated by the limited size of the simulated  samples. Single-event sensitivities, defined as signal branching ratios corresponding to the observation of one signal event, are found to be
\begin{eqnarray}
{\cal B}_{\rm SES}(K^+\to\pi^0\pi^-\mu^+e^+) =
1/\! \left(N_K \cdot {\cal B}_{\gamma\gamma}
\cdot A(K^+\to\pi^0\pi^-\mu^+e^+) \right) &
\!\!\!=\!\!\! & (1.26\pm0.05)\times 10^{-10}, \nonumber \\
{\cal B}_{\rm SES}(K^+\to\pi^0\pi^+\mu^-e^+) =
1/\! \left(N_K \cdot {\cal B}_{\gamma\gamma}
\cdot A(K^+\to\pi^0\pi^+\mu^-e^+) \right) &
\!\!\!=\!\!\! & (1.37\pm0.05)\times 10^{-10}, \nonumber \\
{\cal B}_{\rm SES}(K^+\to\pi^0\pi^+\mu^+e^-) =
1/\! \left(N_K \cdot {\cal B}_{\gamma\gamma}
\cdot A(K^+\to\pi^0\pi^+\mu^+e^-) \right) &
\!\!\!=\!\!\! & (2.17\pm0.09)\times 10^{-10}. \nonumber
\end{eqnarray}
Here $N_K=(1.97\pm0.07)\times 10^{12}$ is the effective number of kaon decays in the FV (Section~\ref{sec:piee}), and ${\cal B}_{\gamma\gamma}=(98.823 \pm 0.034)\%$ is the $\pi^0\to\gamma\gamma$ branching ratio~\cite{pdg}. The uncertainties in ${\cal B}_{\rm SES}$ are
%due to those in $N_K$ (
dominated by those in the external parameter ${\cal B}_{\pi ee}$ and in the signal acceptances.

After unmasking the signal mass regions, no events are observed in the data for any of the three signal modes (Fig.~\ref{fig:signal}, right). Upper limits of the signal branching ratios at 90\% CL are evaluated using the CL$_{\rm S}$ method~\cite{re02}:
\begin{eqnarray}
{\cal B}(K^+\to\pi^0\pi^-\mu^+e^+) & < & 2.9\times 10^{-10}, \nonumber \\
{\cal B}(K^+\to\pi^0\pi^+\mu^-e^+) & < & 3.1\times 10^{-10}, \nonumber \\
{\cal B}(K^+\to\pi^0\pi^+\mu^+e^-) & < & 5.0\times 10^{-10}. \nonumber
\end{eqnarray}

%%%%%%%%%%%%%%%%%%%%%%%%%%%%

\section*{Summary}

The first search for the lepton number violating decay $K^+\to\pi^0\pi^-\mu^+e^+$ and lepton flavour violating decays $K^+\to\pi^0\pi^+\mu^-e^+$, $K^+\to\pi^0\pi^+\mu^+e^-$ has been performed using the di-lepton dataset collected by the NA62 experiment at CERN in 2016--2018. Upper limits of $2.9\times 10^{-10}$, $3.1\times 10^{-10}$ and $5.0\times 10^{-10}$, respectively, are obtained at 90\% CL for the branching ratios of the three decays on the assumption of uniform phase-space distributions.

%%%%%%%%%%%%%%%%%%%%%%%%%%%%%%%%%%%%%%%%%%%%

\section*{Acknowledgements}

\input{acknow}

%\end{linenumbers}

%%%%%%%%%%%%%%%%%%%%%%%%%%%%%%%%%%%%%%%%%%%%

\newpage
\input{authors}

\clearpage

\end{document}

%% file: acknow.tex
It is a pleasure to express our appreciation to the staff of the CERN laboratory and the technical
staff of the participating laboratories and universities for their efforts in the operation of the
experiment and data processing.

The cost of the experiment and its auxiliary systems was supported by the funding agencies of 
the Collaboration Institutes. We are particularly indebted to: 
F.R.S.-FNRS (Fonds de la Recherche Scientifique - FNRS), under Grants No. 4.4512.10, 1.B.258.20, Belgium;
CECI (Consortium des Equipements de Calcul Intensif), funded by the Fonds de la Recherche Scientifique de Belgique (F.R.S.-FNRS) under Grant No. 2.5020.11 and by the Walloon Region, Belgium;
%BMES (Ministry of Education, Youth and Science), Bulgaria;
NSERC (Natural Sciences and Engineering Research Council), funding SAPPJ-2018-0017,  Canada;
%NRC (National Research Council) contribution to TRIUMF, Canada;
MEYS (Ministry of Education, Youth and Sports) funding LM 2018104, Czech Republic;
BMBF (Bundesministerium f\"{u}r Bildung und Forschung) contracts 05H12UM5, 05H15UMCNA and 05H18UMCNA, Germany;
INFN  (Istituto Nazionale di Fisica Nucleare),  Italy;
MIUR (Ministero dell'Istruzione, dell'Universit\`a e della Ricerca),  Italy;
CONACyT  (Consejo Nacional de Ciencia y Tecnolog\'{i}a),  Mexico;
IFA (Institute of Atomic Physics) Romanian 
% ended in 2019
CERN-RO No. 1/16.03.2016 
% 2020 and 2021
%CERN-RO Nr. 10/10.03.2020
% 2022-2024
%CERN-RO Nr. 06/03.01.2022
and Nucleus Programme PN 19 06 01 04,  Romania;
% remove for opt C
%INR-RAS (Institute for Nuclear Research of the Russian Academy of Sciences), Moscow, Russia; 
%JINR (Joint Institute for Nuclear Research), Dubna, Russia; 
%NRC (National Research Center)  ``Kurchatov Institute'' and MESRF (Ministry of Education and Science of the Russian Federation), Russia; 
MESRS  (Ministry of Education, Science, Research and Sport), Slovakia; 
CERN (European Organization for Nuclear Research), Switzerland; 
STFC (Science and Technology Facilities Council), United Kingdom;
NSF (National Science Foundation) Award Numbers 1506088 and 1806430,  U.S.A.;
ERC (European Research Council)  ``UniversaLepto'' advanced grant 268062, ``KaonLepton'' starting grant 336581, Europe.

Individuals have received support from:
Charles University (Research Center UNCE/SCI/013, grant PRIMUS 23/SCI/025), Czech Republic;
Czech Science Foundation (grant 23-06770S);
Ministero dell'Istruzione, dell'Universit\`a e della Ricerca (MIUR  ``Futuro in ricerca 2012''  grant RBFR12JF2Z, Project GAP), Italy;
% terminnated
% Russian Foundation for Basic Research  (RFBR grants 18-32-00072, 18-32-00245), Russia; 
% remove for opt C
%Russian Science Foundation (RSF 19-72-10096), Russia;
the Royal Society  (grants UF100308, UF0758946), United Kingdom;
STFC (Rutherford fellowships ST/J00412X/1, ST/M005798/1), United Kingdom;
ERC (grants 268062,  336581 and  starting grant 802836 ``AxScale'');
EU Horizon 2020 (Marie Sk\l{}odowska-Curie grants 701386, 754496, 842407, 893101, 101023808).
% remove for opt C
%The data used in this paper were collected before February 2022.

%% file: authors.tex
\newcommand{\orcimg}{\raisebox{-0.3\height}{\includegraphics[height=\fontcharht\font`A]{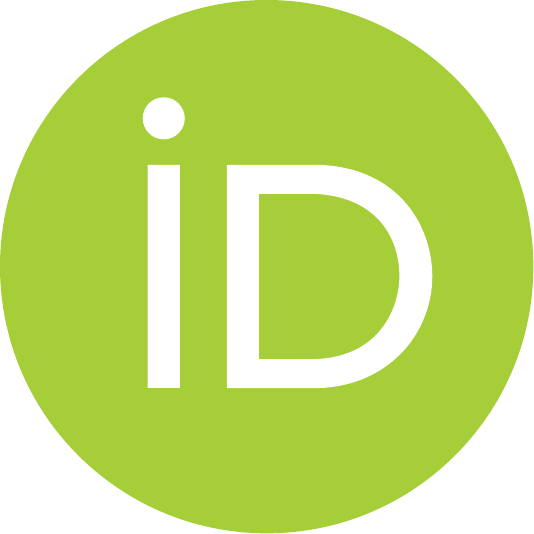}}}
\newcommand{\orcid}[1]{\href{https://orcid.org/#1}{\orcimg}}

\centerline{\bf The NA62 Collaboration} 
\vspace{0.7cm}
%
%%%%%%%%%%%%%%%%%%%%%%%%%%%%%%%%%
%

\begin{raggedright}
\noindent
%%%%%%%
{\bf Universit\'e Catholique de Louvain, Louvain-La-Neuve, Belgium}\\
 M. U.~Ashraf\orcid{0000-0001-8855-8348},
 E.~Cortina Gil\orcid{0000-0001-9627-699X},
 E.~Minucci$\,${\footnotemark[1]}\orcid{0000-0002-3972-6824},
 S.~Padolski\orcid{0000-0002-6795-7670},
 P.~Petrov,
 A.~Shaikhiev$\,${\footnotemark[2]}\orcid{0000-0003-2921-8743},
 R.~Volpe$\,${\footnotemark[3]}\orcid{0000-0003-1782-2978}
\vspace{0.4cm}

%%%%%%%
{\bf TRIUMF, Vancouver, British Columbia, Canada}\\
 T.~Numao\orcid{0000-0001-5232-6190},
 Y.~Petrov\orcid{0000-0003-2643-8740},
 B.~Velghe\orcid{0000-0002-0797-8381},
 V. W. S.~Wong\orcid{0000-0001-5975-8164}
\vspace{0.4cm}

%%%%%%%
{\bf University of British Columbia, Vancouver, British Columbia, Canada}\\
 D.~Bryman$\,${\footnotemark[4]}\orcid{0000-0002-9691-0775},
 J.~Fu
\vspace{0.4cm}

%%%%%%%
{\bf Charles University, Prague, Czech Republic}\\
 Z.~Hives\orcid{0000-0002-5025-993X},
 T.~Husek$\,${\footnotemark[5]}\orcid{0000-0002-7208-9150},
 J.~Jerhot$\,${\footnotemark[6]}\orcid{0000-0002-3236-1471},
 K.~Kampf\orcid{0000-0003-1096-667X},
 M.~Kolesar\orcid{0000-0002-9085-2252},
 M.~Zamkovsky$\,${\footnotemark[7]}\orcid{0000-0002-5067-4789}
\vspace{0.4cm}

%%%%%%%
{\bf Aix Marseille University, CNRS/IN2P3, CPPM, Marseille, France}\\
 B.~De Martino\orcid{0000-0003-2028-9326},
 M.~Perrin-Terrin\orcid{0000-0002-3568-1956}
\vspace{0.4cm}

%%%%%%%
{\bf Max-Planck-Institut f\"ur Physik (Werner-Heisenberg-Institut), Garching, Germany}\\
 B.~D\"obrich\orcid{0000-0002-6008-8601},
 S.~Lezki\orcid{0000-0002-6909-774X},
 J.~Schubert$\,${\footnotemark[8]}\orcid{0000-0002-5782-8816}
\vspace{0.4cm}

%%%%%%%
{\bf Institut f\"ur Physik and PRISMA Cluster of Excellence, Universit\"at Mainz, Mainz, Germany}\\
 A. T.~Akmete\orcid{0000-0002-5580-5477},
 R.~Aliberti$\,${\footnotemark[9]}\orcid{0000-0003-3500-4012},
 G.~Khoriauli$\,${\footnotemark[10]}\orcid{0000-0002-6353-8452},
 J.~Kunze,
 D.~Lomidze$\,${\footnotemark[11]}\orcid{0000-0003-3936-6942}, 
 L.~Peruzzo\orcid{0000-0002-4752-6160},
 M.~Vormstein,
 R.~Wanke\orcid{0000-0002-3636-360X}
\vspace{0.4cm}

%%%%%%%
{\bf Dipartimento di Fisica e Scienze della Terra dell'Universit\`a e INFN, Sezione di Ferrara, Ferrara, Italy}\\
 P.~Dalpiaz,
 M.~Fiorini\orcid{0000-0001-6559-2084},
 I.~Neri\orcid{0000-0002-9669-1058},
 A.~Norton$\,${\footnotemark[12]}\orcid{0000-0001-5959-5879},
 F.~Petrucci\orcid{0000-0002-7220-6919}, 
 M.~Soldani\orcid{0000-0003-4902-943X},
 H.~Wahl$\,${\footnotemark[13]}\orcid{0000-0003-0354-2465}
\vspace{0.4cm}

%%%%%%%
{\bf INFN, Sezione di Ferrara, Ferrara, Italy}\\
 L.~Bandiera\orcid{0000-0002-5537-9674},
 A.~Cotta Ramusino\orcid{0000-0003-1727-2478},
 A.~Gianoli\orcid{0000-0002-2456-8667},
 M.~Romagnoni\orcid{0000-0002-2775-6903},
 A.~Sytov\orcid{0000-0001-8789-2440}
\vspace{0.4cm}

%%%%%%%
{\bf Dipartimento di Fisica e Astronomia dell'Universit\`a e INFN, Sezione di Firenze, Sesto Fiorentino, Italy}\\
 E.~Iacopini\orcid{0000-0002-5605-2497},
 G.~Latino\orcid{0000-0002-4098-3502},
 M.~Lenti\orcid{0000-0002-2765-3955},
 P.~Lo Chiatto\orcid{0000-0002-4177-557X},
 I.~Panichi\orcid{0000-0001-7749-7914},
 A.~Parenti\orcid{0000-0002-6132-5680}
\vspace{0.4cm}

%%%%%%%
{\bf INFN, Sezione di Firenze, Sesto Fiorentino, Italy}\\
 A.~Bizzeti$\,${\footnotemark[14]}\orcid{0000-0001-5729-5530},
 F.~Bucci\orcid{0000-0003-1726-3838}
\vspace{0.4cm}

%%%%%%%
{\bf Laboratori Nazionali di Frascati, Frascati, Italy}\\
 A.~Antonelli\orcid{0000-0001-7671-7890},
 G.~Georgiev$\,${\footnotemark[15]}\orcid{0000-0001-6884-3942},
 V.~Kozhuharov$\,${\footnotemark[15]}\orcid{0000-0002-0669-7799},
 G.~Lanfranchi\orcid{0000-0002-9467-8001},
 S.~Martellotti\orcid{0000-0002-4363-7816}, 
 M.~Moulson\orcid{0000-0002-3951-4389},
 T.~Spadaro\orcid{0000-0002-7101-2389},
 G.~Tinti\orcid{0000-0003-1364-844X}
\vspace{0.4cm}

%%%%%%%
{\bf Dipartimento di Fisica ``Ettore Pancini'' e INFN, Sezione di Napoli, Napoli, Italy}\\
 F.~Ambrosino\orcid{0000-0001-5577-1820},
 T.~Capussela,
 M.~Corvino\orcid{0000-0002-2401-412X},
 M.~D'Errico\orcid{0000-0001-5326-1106},
 D.~Di Filippo\orcid{0000-0003-1567-6786}, 
 R.~Fiorenza$\,${\footnotemark[16]}\orcid{0000-0003-4965-7073},
 M.~Francesconi\orcid{0000-0002-7029-7634},
 R.~Giordano\orcid{0000-0002-5496-7247},
 P.~Massarotti\orcid{0000-0002-9335-9690},
 M.~Mirra\orcid{0000-0002-1190-2961},
 M.~Napolitano\orcid{0000-0003-1074-9552}, 
 I.~Rosa\orcid{0009-0002-7564-1825},
 G.~Saracino\orcid{0000-0002-0714-5777}
\vspace{0.4cm}

%%%%%%%
{\bf Dipartimento di Fisica e Geologia dell'Universit\`a e INFN, Sezione di Perugia, Perugia, Italy}\\
 G.~Anzivino\orcid{0000-0002-5967-0952},
 F.~Brizioli$\,${\footnotemark[7]}\orcid{0000-0002-2047-441X},
 E.~Imbergamo,
 R.~Lollini\orcid{0000-0003-3898-7464},
 R.~Piandani$\,${\footnotemark[17]}\orcid{0000-0003-2226-8924},
 C.~Santoni\orcid{0000-0001-7023-7116}
%\vspace{0.5cm}
\newpage

%%%%%%%
{\bf INFN, Sezione di Perugia, Perugia, Italy}\\
 M.~Barbanera\orcid{0000-0002-3616-3341},
 P.~Cenci\orcid{0000-0001-6149-2676},
 B.~Checcucci\orcid{0000-0002-6464-1099},
 P.~Lubrano\orcid{0000-0003-0221-4806},
 M.~Lupi$\,${\footnotemark[18]}\orcid{0000-0001-9770-6197}, 
 M.~Pepe\orcid{0000-0001-5624-4010},
 M.~Piccini\orcid{0000-0001-8659-4409}
\vspace{0.5cm}

%%%%%%%
{\bf Dipartimento di Fisica dell'Universit\`a e INFN, Sezione di Pisa, Pisa, Italy}\\
 F.~Costantini\orcid{0000-0002-2974-0067},
 L.~Di Lella$\,${\footnotemark[13]}\orcid{0000-0003-3697-1098},
 N.~Doble$\,${\footnotemark[13]}\orcid{0000-0002-0174-5608},
 M.~Giorgi\orcid{0000-0001-9571-6260},
 S.~Giudici\orcid{0000-0003-3423-7981}, 
 G.~Lamanna\orcid{0000-0001-7452-8498},
 E.~Lari\orcid{0000-0003-3303-0524},
 E.~Pedreschi\orcid{0000-0001-7631-3933},
 M.~Sozzi\orcid{0000-0002-2923-1465}
\vspace{0.5cm}

%%%%%%%
{\bf INFN, Sezione di Pisa, Pisa, Italy}\\
 C.~Cerri,
 R.~Fantechi\orcid{0000-0002-6243-5726},
 L.~Pontisso$\,${\footnotemark[19]}\orcid{0000-0001-7137-5254},
 F.~Spinella\orcid{0000-0002-9607-7920}
\vspace{0.5cm}

%%%%%%%
{\bf Scuola Normale Superiore e INFN, Sezione di Pisa, Pisa, Italy}\\
 I.~Mannelli\orcid{0000-0003-0445-7422}
\vspace{0.5cm}

%%%%%%%
{\bf Dipartimento di Fisica, Sapienza Universit\`a di Roma e INFN, Sezione di Roma I, Roma, Italy}\\
 G.~D'Agostini\orcid{0000-0002-6245-875X},
 M.~Raggi\orcid{0000-0002-7448-9481}
\vspace{0.5cm}

%%%%%%%
{\bf INFN, Sezione di Roma I, Roma, Italy}\\
 A.~Biagioni\orcid{0000-0001-5820-1209},
 P.~Cretaro\orcid{0000-0002-2229-149X},
 O.~Frezza\orcid{0000-0001-8277-1877},
 E.~Leonardi\orcid{0000-0001-8728-7582},
 A.~Lonardo\orcid{0000-0002-5909-6508}, 
 M.~Turisini\orcid{0000-0002-5422-1891},
 P.~Valente\orcid{0000-0002-5413-0068},
 P.~Vicini\orcid{0000-0002-4379-4563}
\vspace{0.5cm}

%%%%%%%
{\bf INFN, Sezione di Roma Tor Vergata, Roma, Italy}\\
 R.~Ammendola\orcid{0000-0003-4501-3289},
 V.~Bonaiuto$\,${\footnotemark[20]}\orcid{0000-0002-2328-4793},
 A.~Fucci,
 A.~Salamon\orcid{0000-0002-8438-8983},
 F.~Sargeni$\,${\footnotemark[21]}\orcid{0000-0002-0131-236X}
\vspace{0.5cm}

%%%%%%%
{\bf Dipartimento di Fisica dell'Universit\`a e INFN, Sezione di Torino, Torino, Italy}\\
 R.~Arcidiacono$\,${\footnotemark[22]}\orcid{0000-0001-5904-142X},
 B.~Bloch-Devaux$\,${\footnotemark[5]}$^,$$\,${\footnotemark[23]}\orcid{0000-0002-2463-1232},
 M.~Boretto$\,${\footnotemark[7]}\orcid{0000-0001-5012-4480},
 E.~Menichetti\orcid{0000-0001-7143-8200},
 E.~Migliore\orcid{0000-0002-2271-5192},
 D.~Soldi\orcid{0000-0001-9059-4831}
\vspace{0.5cm}

%%%%%%%
{\bf INFN, Sezione di Torino, Torino, Italy}\\
 C.~Biino$\,${\footnotemark[24]}\orcid{0000-0002-1397-7246},
 A.~Filippi\orcid{0000-0003-4715-8748},
 F.~Marchetto\orcid{0000-0002-5623-8494}
\vspace{0.5cm}

%%%%%%%
{\bf Instituto de F\'isica, Universidad Aut\'onoma de San Luis Potos\'i, San Luis Potos\'i, Mexico}\\
 A.~Briano Olvera\orcid{0000-0001-6121-3905},
 J.~Engelfried\orcid{0000-0001-5478-0602},
 N.~Estrada-Tristan$\,${\footnotemark[25]}\orcid{0000-0003-2977-9380},
 M.~A.~Reyes Santos$\,${\footnotemark[25]}\orcid{0000-0003-1347-2579},
 K.~A.~Rodriguez Rivera\orcid{0000-0001-5723-9176}
\vspace{0.5cm}

%%%%%%%
{\bf Horia Hulubei National Institute for R\&D in Physics and Nuclear Engineering, Bucharest-Magurele, Romania}\\
 P.~Boboc\orcid{0000-0001-5532-4887},
 A. M.~Bragadireanu,
 S. A.~Ghinescu\orcid{0000-0003-3716-9857},
 O. E.~Hutanu
\vspace{0.5cm}

%%%%%%%
{\bf Faculty of Mathematics, Physics and Informatics, Comenius University, Bratislava, Slovakia}\\
 L.~Bician$\,${\footnotemark[26]}\orcid{0000-0001-9318-0116},
 T.~Blazek\orcid{0000-0002-2645-0283},
 V.~Cerny\orcid{0000-0003-1998-3441},
 Z.~Kucerova$\,${\footnotemark[7]}\orcid{0000-0001-8906-3902},
 T.~Velas\orcid{0009-0004-0061-1968}
\vspace{0.5cm}

%%%%%%%
{\bf CERN, European Organization for Nuclear Research, Geneva, Switzerland}\\
 J.~Bernhard\orcid{0000-0001-9256-971X},
 A.~Ceccucci\orcid{0000-0002-9506-866X},
 M.~Ceoletta\orcid{0000-0002-2532-0217},
 H.~Danielsson\orcid{0000-0002-1016-5576},
 N.~De Simone$\,${\footnotemark[27]}, 
 F.~Duval,
 L.~Federici\orcid{0000-0002-3401-9522},
 E.~Gamberini\orcid{0000-0002-6040-4985},
 L.~Gatignon$\,${\footnotemark[2]}\orcid{0000-0001-6439-2945},
 R.~Guida, 
 F.~Hahn$\,$\renewcommand{\thefootnote}{\fnsymbol{footnote}}\footnotemark[2]\renewcommand{\thefootnote}{\arabic{footnote}},
 E.~B.~Holzer\orcid{0000-0003-2622-6844},
 B.~Jenninger,
 M.~Koval$\,${\footnotemark[26]}\orcid{0000-0002-6027-317X},
 P.~Laycock$\,${\footnotemark[28]}\orcid{0000-0002-8572-5339}, 
 G.~Lehmann Miotto\orcid{0000-0001-9045-7853},
 P.~Lichard\orcid{0000-0003-2223-9373},
 A.~Mapelli\orcid{0000-0002-4128-1019},
 K.~Massri$\,${\footnotemark[2]}\orcid{0000-0001-7533-6295},
 M.~Noy, 
 V.~Palladino\orcid{0000-0002-9786-9620},
 J.~Pinzino$\,${\footnotemark[29]}\orcid{0000-0002-7418-0636},
 V.~Ryjov,
 S.~Schuchmann\orcid{0000-0002-8088-4226},
 S.~Venditti
\vspace{0.5cm}

%%%%%%%
{\bf Ecole Polytechnique F\'ed\'erale Lausanne, Lausanne, Switzerland}\\
 X.~Chang\orcid{0000-0002-8792-928X},
 A.~Kleimenova\orcid{0000-0002-9129-4985},
 R.~Marchevski\orcid{0000-0003-3410-0918}
\vspace{0.5cm}

%%%%%%%
{\bf School of Physics and Astronomy, University of Birmingham, Birmingham, United Kingdom}\\
 T.~Bache\orcid{0000-0003-4520-830X},
 M. B.~Brunetti$\,${\footnotemark[30]}\orcid{0000-0003-1639-3577},
 V.~Duk$\,${\footnotemark[3]}\orcid{0000-0001-6440-0087},
 V.~Fascianelli$\,${\footnotemark[31]},
 J. R.~Fry\orcid{0000-0002-3680-361X}, 
 F.~Gonnella\orcid{0000-0003-0885-1654},
 E.~Goudzovski$\,$\renewcommand{\thefootnote}{\fnsymbol{footnote}}\footnotemark[1]\renewcommand{\thefootnote}{\arabic{footnote}}\orcid{0000-0001-9398-4237},
 J.~Henshaw\orcid{0000-0001-7059-421X},
 L.~Iacobuzio,
 C.~Kenworthy\orcid{0009-0002-8815-0048}, 
 C.~Lazzeroni\orcid{0000-0003-4074-4787},
 N.~Lurkin\orcid{0000-0002-9440-5927},
 F.~Newson,
 C.~Parkinson\orcid{0000-0003-0344-7361},
 A.~Romano\orcid{0000-0003-1779-9122}, 
 J.~Sanders\orcid{0000-0003-1014-094X},
 A.~Sergi$\,${\footnotemark[32]}\orcid{0000-0001-9495-6115},
 A.~Sturgess\orcid{0000-0002-8104-5571},
 J.~Swallow$\,${\footnotemark[33]}\orcid{0000-0002-1521-0911},
 A.~Tomczak\orcid{0000-0001-5635-3567}
\vspace{0.5cm}

%%%%%%%
{\bf School of Physics, University of Bristol, Bristol, United Kingdom}\\
 H.~Heath\orcid{0000-0001-6576-9740},
 R.~Page,
 S.~Trilov\orcid{0000-0003-0267-6402}
\vspace{0.5cm}

%%%%%%%
{\bf School of Physics and Astronomy, University of Glasgow, Glasgow, United Kingdom}\\
 B.~Angelucci,
 D.~Britton\orcid{0000-0001-9998-4342},
 C.~Graham\orcid{0000-0001-9121-460X},
 D.~Protopopescu\orcid{0000-0002-8047-6513}
\vspace{0.5cm}

%%%%%%%
{\bf Physics Department, University of Lancaster, Lancaster, United Kingdom}\\
 J.~Carmignani$\,${\footnotemark[34]}\orcid{0000-0002-1705-1061},
 J. B.~Dainton,
 R. W. L.~Jones\orcid{0000-0002-6427-3513},
 G.~Ruggiero$\,${\footnotemark[35]}\orcid{0000-0001-6605-4739}
\vspace{0.5cm}

%%%%%%%
{\bf School of Physical Sciences, University of Liverpool, Liverpool, United Kingdom}\\
 L.~Fulton,
 D.~Hutchcroft\orcid{0000-0002-4174-6509},
 E.~Maurice$\,${\footnotemark[36]}\orcid{0000-0002-7366-4364},
 B.~Wrona\orcid{0000-0002-1555-0262}
\vspace{0.5cm}

%%%%%%%
{\bf Physics and Astronomy Department, George Mason University, Fairfax, Virginia, USA}\\
 A.~Conovaloff,
 P.~Cooper,
 D.~Coward$\,${\footnotemark[37]}\orcid{0000-0001-7588-1779},
 P.~Rubin\orcid{0000-0001-6678-4985}
\vspace{0.5cm}

%%%%%%%
{\bf An Institute or an international laboratory covered by a cooperation agreement with CERN}\\
 A.~Baeva,
 D.~Baigarashev$\,${\footnotemark[38]}\orcid{0000-0001-6101-317X},
 V.~Bautin\orcid{0000-0002-5283-6059},
 D.~Emelyanov,
 T.~Enik\orcid{0000-0002-2761-9730}, 
 V.~Falaleev$\,${\footnotemark[3]}\orcid{0000-0003-3150-2196},
 S.~Fedotov,
 K.~Gorshanov\orcid{0000-0001-7912-5962},
 E.~Gushchin\orcid{0000-0001-8857-1665},
 V.~Kekelidze\orcid{0000-0001-8122-5065}, 
 D.~Kereibay,
 S.~Kholodenko$\,${\footnotemark[29]}\orcid{0000-0002-0260-6570},
 A.~Khotyantsev,
 A.~Korotkova,
 Y.~Kudenko\orcid{0000-0003-3204-9426}, 
 V.~Kurochka,
 V.~Kurshetsov\orcid{0000-0003-0174-7336},
 L.~Litov$\,${\footnotemark[15]}\orcid{0000-0002-8511-6883},
 D.~Madigozhin\orcid{0000-0001-8524-3455},
 M.~Medvedeva, 
 A.~Mefodev,
 M.~Misheva$\,${\footnotemark[39]},
 N.~Molokanova,
 S.~Movchan,
 V.~Obraztsov\orcid{0000-0002-0994-3641}, 
 A.~Okhotnikov\orcid{0000-0003-1404-3522},
 A.~Ostankov$\,$\renewcommand{\thefootnote}{\fnsymbol{footnote}}\footnotemark[2]\renewcommand{\thefootnote}{\arabic{footnote}},
 I.~Polenkevich,
 Yu.~Potrebenikov\orcid{0000-0003-1437-4129},
 A.~Sadovskiy\orcid{0000-0002-4448-6845}, 
 K.~Salamatin\orcid{0000-0001-6287-8685},
 V.~Semenov$\,$\renewcommand{\thefootnote}{\fnsymbol{footnote}}\footnotemark[2]\renewcommand{\thefootnote}{\arabic{footnote}},
 S.~Shkarovskiy,
 V.~Sugonyaev\orcid{0000-0003-4449-9993},
 O.~Yushchenko\orcid{0000-0003-4236-5115}, 
 A.~Zinchenko$\,$\renewcommand{\thefootnote}{\fnsymbol{footnote}}\footnotemark[2]\renewcommand{\thefootnote}{\arabic{footnote}}
\vspace{0.5cm}

\end{raggedright}

%
%%%%%%%%%%%%%%%%%%%%%%%%%%%%%%%%%
%

\setcounter{footnote}{0}
\newlength{\basefootnotesep}
\setlength{\basefootnotesep}{\footnotesep}

\renewcommand{\thefootnote}{\fnsymbol{footnote}}
\noindent
$^{\footnotemark[1]}${Corresponding author: E.~Goudzovski, email: evgueni.goudzovski@cern.ch}\\
$^{\footnotemark[2]}${Deceased}\\
\renewcommand{\thefootnote}{\arabic{footnote}}
$^{1}${Present address: Syracuse University, Syracuse, NY 13244, USA} \\
$^{2}${Present address: Physics Department, University of Lancaster, Lancaster, LA1 4YB, UK} \\
$^{3}${Present address: INFN, Sezione di Perugia, I-06100 Perugia, Italy} \\
$^{4}${Also at TRIUMF, Vancouver, British Columbia, V6T 2A3, Canada} \\
$^{5}${Also at School of Physics and Astronomy, University of Birmingham, Birmingham, B15 2TT, UK} \\
$^{6}${Present address: Max-Planck-Institut f\"ur Physik (Werner-Heisenberg-Institut), Garching, D-85748, Germany} \\
$^{7}${Present address: CERN, European Organization for Nuclear Research, CH-1211 Geneva 23, Switzerland} \\
$^{8}${Also at Department of Physics, Technical University of Munich, M\"unchen, D-80333, Germany} \\
$^{9}${Present address: Institut f\"ur Kernphysik and Helmholtz Institute Mainz, Universit\"at Mainz, Mainz, D-55099, Germany} \\
$^{10}${Present address: Universit\"at W\"urzburg, D-97070 W\"urzburg, Germany} \\
$^{11}${Present address: European XFEL GmbH, D-22869 Schenefeld, Germany} \\
$^{12}${Present address: School of Physics and Astronomy, University of Glasgow, Glasgow, G12 8QQ, UK} \\
$^{13}${Present address: Institut f\"ur Physik and PRISMA Cluster of Excellence, Universit\"at Mainz, D-55099 Mainz, Germany} \\
$^{14}${Also at Dipartimento di Scienze Fisiche, Informatiche e Matematiche, Universit\`a di Modena e Reggio Emilia, I-41125 Modena, Italy} \\
$^{15}${Also at Faculty of Physics, University of Sofia, BG-1164 Sofia, Bulgaria} \\
$^{16}${Present address: Scuola Superiore Meridionale e INFN, Sezione di Napoli, I-80138 Napoli, Italy} \\
$^{17}${Present address: Instituto de F\'isica, Universidad Aut\'onoma de San Luis Potos\'i, 78240 San Luis Potos\'i, Mexico} \\
$^{18}${Present address: Institut am Fachbereich Informatik und Mathematik, Goethe Universit\"at, D-60323 Frankfurt am Main, Germany} \\
$^{19}${Present address: INFN, Sezione di Roma I, I-00185 Roma, Italy} \\
$^{20}${Also at Department of Industrial Engineering, University of Roma Tor Vergata, I-00173 Roma, Italy} \\
$^{21}${Also at Department of Electronic Engineering, University of Roma Tor Vergata, I-00173 Roma, Italy} \\
$^{22}${Also at Universit\`a degli Studi del Piemonte Orientale, I-13100 Vercelli, Italy} \\
$^{23}${Present address: Universit\'e Catholique de Louvain, B-1348 Louvain-La-Neuve, Belgium} \\
$^{24}${Also at Gran Sasso Science Institute, I-67100 L'Aquila,  Italy} \\
$^{25}${Also at Universidad de Guanajuato, 36000 Guanajuato, Mexico} \\
$^{26}${Present address: Charles University, 116 36 Prague 1, Czech Republic} \\
$^{27}${Present address: DESY, D-15738 Zeuthen, Germany} \\
$^{28}${Present address: Brookhaven National Laboratory, Upton, NY 11973, USA} \\
$^{29}${Present address: INFN, Sezione di Pisa, I-56100 Pisa, Italy} \\
$^{30}${Present address: Department of Physics, University of Warwick, Coventry, CV4 7AL, UK} \\
$^{31}${Present address: Center for theoretical neuroscience, Columbia University, New York, NY 10027, USA} \\
$^{32}${Present address: Dipartimento di Fisica dell'Universit\`a e INFN, Sezione di Genova, I-16146 Genova, Italy} \\
$^{33}${Present address: Laboratori Nazionali di Frascati, I-00044 Frascati, Italy} \\
$^{34}${Present address: School of Physical Sciences, University of Liverpool, Liverpool, L69 7ZE, UK} \\
$^{35}${Present address: Dipartimento di Fisica e Astronomia dell'Universit\`a e INFN, Sezione di Firenze, I-50019 Sesto Fiorentino, Italy} \\
$^{36}${Present address: Laboratoire Leprince Ringuet, F-91120 Palaiseau, France} \\
$^{37}${Also at SLAC National Accelerator Laboratory, Stanford University, Menlo Park, CA 94025, USA} \\
$^{38}${Also at L. N. Gumilyov Eurasian National University, 010000 Nur-Sultan, Kazakhstan} \\
$^{39}${Present address: Institute of Nuclear Research and Nuclear Energy of Bulgarian Academy of Science (INRNE-BAS), BG-1784 Sofia, Bulgaria} \\